\documentclass[10pt,twocolumn]{article}  
\usepackage{epsfig,graphicx,graphics,amsmath}  
\usepackage[comma,authoryear]{natbib}  
\usepackage{fancyhdr}  
\input epsf  
  
\newcommand{\beq}{\begin{equation}}  
\newcommand{\eeq}{\end{equation}}  
\newcommand{\ben}{\begin{enumerate}}
\newcommand{\een}{\end{enumerate}}
\newcommand{\bitem}{\begin{itemize}}
\newcommand{\eitem}{\end{itemize}}
\newcommand{\bverb}{\begin{verbatim}}
\newcommand{\everb}{\end{verbatim}}
\newcommand{\bfig}{\begin{figure}}
\newcommand{\efig}{\end{figure}}
\newcommand{\bcen}{\begin{center}}
\newcommand{\ecen}{\end{center}}

\newcommand{\delete}[1]{}

\title{\bf Deficient Reasoning for Dark Matter in Galaxies}
\author{James Q. Feng and C. F. Gallo \\
Superconix Inc. 2440 Lisbon Avenue, Lake Elmo, MN 55042, USA \\
james.q.feng@gmail.com}

\begin{document}
\fancypagestyle{newstyle}{
\fancyhf{}
\fancyhead[C]{James Q. Feng and C. F. Gallo}
\fancyfoot[C]{\thepage}
}
\pagestyle{newstyle}

\twocolumn[
\begin{@twocolumnfalse}
\maketitle


\hrule

\begin{abstract}
In this universe, not all of the matter around us can be readily seen.
The further an object is away from us and the less luminous it is,
the less visible it becomes.
Just by looking at an object is usually difficult,
if not impossible, to tell the amount of mass it contains.
But astronomers have been using the measured luminosity 
to estimate the {\em luminous mass} of stars,
based on empirically established mass-to-light ratio 
which seems to be only applicable to a special class of stars---the 
main-sequence stars---with still considerable uncertainties.
Another basic tool for astronomers to determine the mass of 
a system of stars or galaxies comes from the study of their motion,
as Newton demonstrated with his law of gravitation,
which yields the {\em gravitational mass}.
Because the luminous mass can at best only represent a 
portion of the gravitational mass,
finding the luminous mass to be different or less than the 
gravitational mass should not be surprising.
Using such an apparent discrepancy as compelling evidence 
for the so-called dark matter, which has been believed to 
possess mysterious nonbaryonic properties having a dominant amount 
in galaxies and the universe, 
seems to be too far a stretch
when seriously examining the facts and uncertainties in
the measurement techniques.
In our opinion, a galaxy with star type distribution varying
from its center to edge may have a mass-to-light ratio varying
accordingly.  With the thin-disk model computations
based on measured rotation curves, we found that
most galaxies have a typical mass density profile that 
peaks at the galactic center and decreases rapidly
within $\sim 5\%$ of the cut-off radius, and then 
declines nearly exponentially toward the edge.
The predicted mass density in the Galactic disk is reasonably 
within the reported range of that observed in interstellar medium.
This leads us to believe that ordinary baryonic matter
can be sufficient for supporting the observed galactic rotation curves;
speculation of large amount of non-baryonic matter 
may be based on an ill-conceived discrepancy between gravitational mass
and luminous mass which appears to be unjustified.

\vspace{5 mm}
\end{abstract}

\hrule

\vspace{10 mm}

\end{@twocolumnfalse}
]


\section{The Popular Belief}
In present days,
a large number of people would believe, as they have been told, that
``dark matter'' makes up about 83\% of the universe by mass, 
and is needed to hold the galaxies together.
Probably few of them actually made sufficient efforts in researching 
the validity of the reasons.
In fact, serious in-depth discussion of such reasons and validity of which 
may not even exist in the ``reputable'' scientific literature,
as we have not been able to satisfactorily find so far.
To this situation, 
one might better be reminded that
``Smart is when you believe only half of what you hear.
Brilliant is when you know which half to believe.'' 

When discussing the dark matter, we should start with its definition.
A common description of dark matter suggests that 
it is a type of matter hypothesized to account for 
effects that appear as a result of mass where no such mass can be
{\em seen}.  It neither emits nor absorbs electromagnetic radiation 
(which includes light) at any significant level;
it is matter not reactant to light,
but its existence and properties are inferred from 
its gravitational effects on visible matter, radiation,
and large-scale structure of the universe \citep{trimble1987}. 

The reason for astrophsicists to hypothesize dark matter 
seems to be the discrepancies, as they believe according to their findings,
between the mass of large astronomical objects
determined from their gravitational effects 
and the mass derived
from the luminous matter those objects
contain \citep{freeman2006}.
From how mass is defined in classical physics,
the method for determining mass from its gravitational effect 
is straightforward to comprehend.
Yet, ways of deriving the ``visible'' or ``luminous'' mass 
(from observed
stars, gas, and dust) as usually quoted in literature
do not appear to have been convincingly explained
with scientific rigor, though sometimes can be quite convoluted 
and difficult to follow. 
Apparently people in astrophysics 
tend to rush in with results or ``evidence'' that cannot
always stand up to serious scientific scrutiny.
Nonetheless,
tremendous efforts and resources have been committed to prematurely declared
phenomena that attracted a good deal of press attention.

Here we attempt to examine 
the available evidence for 
such conceived dark matter,
starting with a brief review of 
common methods for measuring mass.
Then, we discuss the resultes from evaluating 
the ``gravitational mass'' and 
``luminous mass'' in galaxies, 
the findings of discrepancy between the two,
and the reasoning for consideration of (mysterious) dark matter as well as 
the difficiencies therewith.
We finally reach our conclusions that
ordinary baryonic matter, some of which may be dark or difficult to see,
could be sufficient for explaining 
the observed galactic rotation, the discrepancy between 
gravitational mass and luminous mass, among other phenomena.

\section{Methods for Determining Mass}
In classical physics, mass is defined as a property of an object 
which determines its resistance to being accelerated by a force 
and the strength of its mutual gravitational attraction with other objects.
As suggested by its definition, 
the (inertial) mass of an object can be determined from
the measured force acting on it and its responsive acceleration,
such as from the ratio of force and acceleration
according to Newton's second law of motion.
This is similar to 
the measurements of other material properties such as 
elasticity 
where the ratio of measured force (or ``stress'') and responsive deformation
(or ``strain'') are used in the calculation.

Weighing an object to determine its mass is a common technique called
{\em gravimetric} method,  one typical form of which is
to use a spring to counteract the force of gravity pulling 
on the object.
In the earth-bound environment, 
the gravimetric method is probably the most precise and reliable
method for measuring mass.
Sophisticated high-precision gravimeters have been used to 
measure density variations in the rocks making up the Earth,
to monitor gravity changes due to mass displacements inside the Earth,  
and to define gravity anomalies.  

When the force is of gravitational nature,
the gravitational field of an object
(which is proportional to its mass) can be determined by measuring
the free-fall acceleration of a small `test object',
and from its gravitational field, the object's 
(active gravitational) mass can be determined.
For example,
a textbook-method for determining the Sun's mass 
\citep{bennett2007}
is to apply the formula of Newton's version of Kepler's 
third law which leads to (by ignoring the Earth's mass comparing to 
that of the Sun) 
\begin{equation} \label{eq:solar-mass}
M_{\odot} \approx \frac{4 \pi^2 \, a^3}{G \, p^2} \approx 2.0 \times 
10^{30} \mbox{ kg} \, \, ,
\end{equation}
where $a$ is the measured average distance between the Earth and the Sun
($\approx 1.5 \times 10^{11}$ m),
$G$ the gravitational constant ($= 6.67 \times 10^{-11}$ 
m$^3$ kg$^{-1}$ s$^{-2}$)\footnote{as can be determined by 
measuring the attraction of two massive objects in a 
sensitive torsion balance},
and $p$ the Earth's orbital period ($\approx 3.15 \times 10^7$ s,
i.e., 1 year).
Actually, (\ref{eq:solar-mass}) can be rearranged as 
\begin{equation} \label{eq:solar-mass2}
\frac{G M_{\odot}}{a^2} \approx \left(\frac{2 \pi \, a}{p}\right)^2 \frac{1}{a} = \frac{V^2}{a} \, \, ,
\end{equation}
where the left side is the gravitational field of the Sun (at 
the Earth's orbit) and 
right side the centripetal acceleration of the Earth 
(with $V$ denoting the magnitude of Earth's orbital velocity).

One of the key variables for determining the mass of a celestial object 
is its distance to a reference position, like $a$ in (\ref{eq:solar-mass}).
For some close objects such as the moon, the planets, the stars in
the local solar neighborhood, 
their distances can be measured by
stellar parallax either from the Earth or by using 
the Earth's orbit. 
Without any assumption about the nature of stars,
parallax is the only technique that 
can determine the distances of stars.
But it is only reliable for accurate measurement of 
stars within a few hundred light years\footnote{in astronomical units (AU),
the distance of a star in parsec (pc) equals the reciprocal 
of the parallax angle in arcseconds 
($1$ pc $= 3.086 \times 10^{16}$ m)}
in the local solar neighborhood
\citep{bennett2007}.

Stellar parallax has enabled measurements of
distances of more than 300 stars within about 10 pc 
(or $\sim 33$ light-years) of the Sun,
among which about half are binary star systems 
consisting of two orbiting stars or multiple star systems 
of three or more stars.
The binary star systems are very important in astrophysics, 
because the information of their orbital motion
provides opportunity to directly determine masses of their 
component stars.
For example, 
the sum of
the two star's masses can be calculated 
from Newton's verison of Kepler's third law 
similar to (\ref{eq:solar-mass}), if both their orbital period
(which is relatively easier to measure) and the distance between them are 
known. 
The individual masses of the two stars can then be determined from
their relative velocities around their common center of mass.
 
Once a star's distance is measured from parallax,
its luminosity can be determined with the inverse square law for 
light intensity. 
With the known luminosity and mass of each individual star for
an observed binary star system, 
an empirical mass-to-light relationship can be determined,
from which the masses of single stars may be estimated indirectly
based on their measured luminosity. 

But in reality, it is often rather difficult to determine the 
average separation between the two stars in a 
binary system,  
because $a$ is needed in using a formula like (\ref{eq:solar-mass}).
Among all types of binary star systems,
only eclipsing binaries of a pair of stars orbiting in
the plane of our line of sight allow detailed study of stellar masses. 

For stars that do not belong to any of the binary star systems
or outside the local solar neighborhood when 
stellar parallax becomes inapplicable, 
their masses may only be estimated indirectly from the established 
value of mass-to-light ratio, which by itself may consist of considerable 
uncertainties.

Stars can have wide ranges of luminosity, surface temperature, and mass.
When measured luminosity and surface temperature of stars 
are plotted in a scatter graph called the Hertzsprung-Russell (H-R) diagram,
some correlative patterns seem to exist between 
the luminosity and surface temperture (or stellar color), especially for 
the so-called ``main-sequence'' stars
that fit in a continuous distinctive band.
All main-sequence stars are fusing hydrogen into helium in their core,
just like the Sun.
Their differences in luminosity and surface temperture 
can be a result of their mass-dependent rate of hydrogen fusion,
because more mass is required to maintain gravitational equilibrium
with the higher rate of nuclear fusion.
Therefore, the mass of a main-sequence star 
may be expected to fall within the same range of 
other stars of the same spectral type 
in the H-R diagram. 
The surface temperature of a star is easier to measure than
luminosity, because it is not expected to change with distance. 
If a star is determined to belong to the main sequence, 
its luminosity may be inferred from the H-R diagram and 
used as a {\em standard candle},
which then enables the calculation of its distance from
its apparent brightness based on the inverse square law---a technique
known as the {\em main-sequence fitting}\footnote{Uncertainty always
exists in main-sequence fitting, become no astonomical object is 
a perfect standard candle; the challenge of finding the objects
that can serve as the best standard candles, therefore, directly relates to
the challenge of measuring astronomical distances}. 
Since its presentation in the first decade of the twentieth century,
the H-R diagram has become one of the most important tools 
in astonomical research, remaining central to the analysis of luminous stars 
\citep{bennett2007}.

However, for objects other than the main-sequence stars,
the reported values of their masses have been derived with 
much less logical clarity and scientific reliability.  
At the present stage of our knowledge, 
``because the methods used for studying the amount of
matter at different scales are so diverse,
there is always the possibility that one or all of the 
estimates could be wrong''
\citep{freeman2006}.

\section{Mass in a Galaxy}
As a stellar system consisting of a large number ($10^5$---$10^{12}$) of 
gravitationally bound stars, an interstellar medium of gas and 
cosmic dust, among others, 
a galaxy has its mass distributed in an extensive space.
Observations have shown that many (late-type, mature) spiral galaxies share 
a common structure with the {\em visible} matter
distributed in a flat thin disk,
rotating about their center of mass 
in nearly circular orbits (cf. Figure 1). 
Many astrophysical systems, such as spiral galaxies, 
planetary systems, planetary rings, accretion disks, etc., appear flat
for a basic reason: 
the state of lowest energy is a flat disk 
perpendicular to an axis along which a distribution of angular momentum
is given for a system of constant mass \citep{binney2008}. 

\begin{figure}
{\includegraphics[clip=true,scale=0.30,viewport=70 60 820 500]{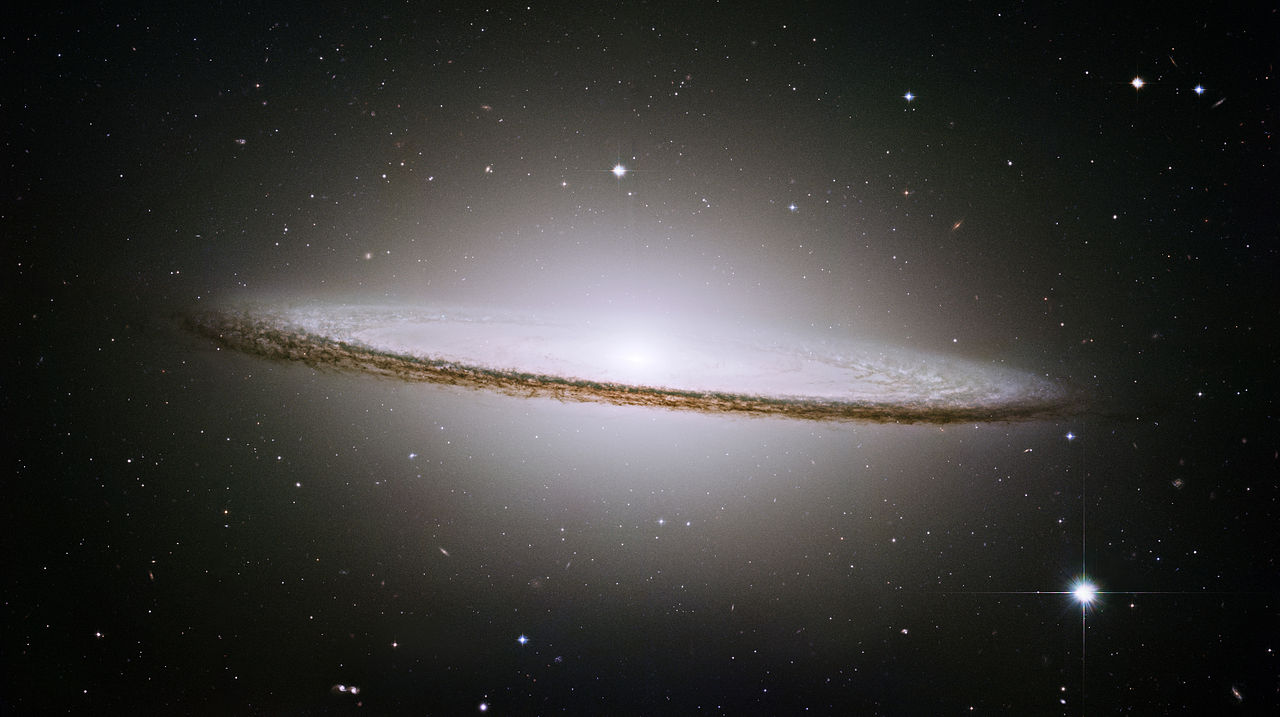}}
{\includegraphics[clip=true,scale=0.080,viewport=100 380 2900 2100]{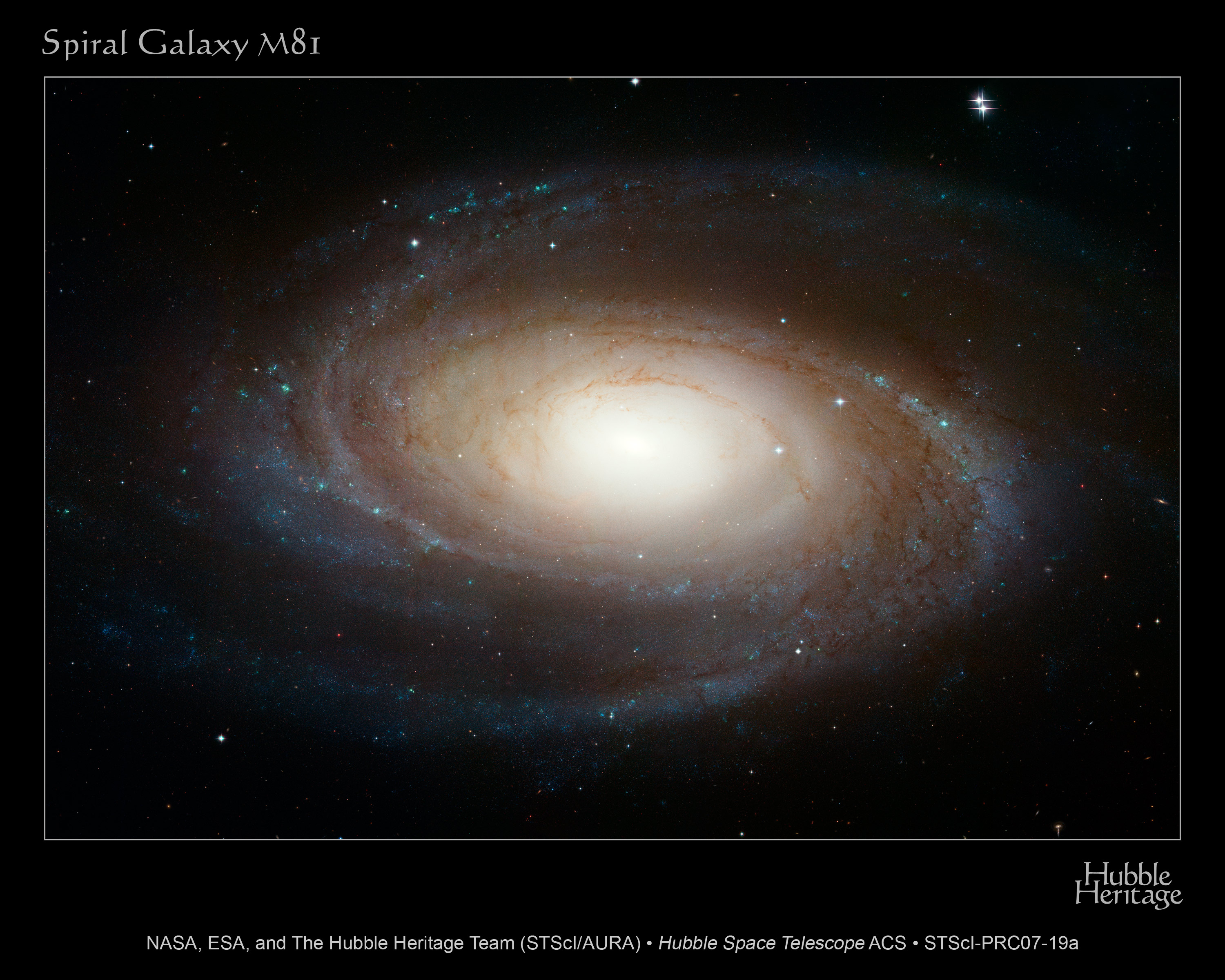}}
{\includegraphics[clip=true,scale=0.472,viewport=50 10 850 300]{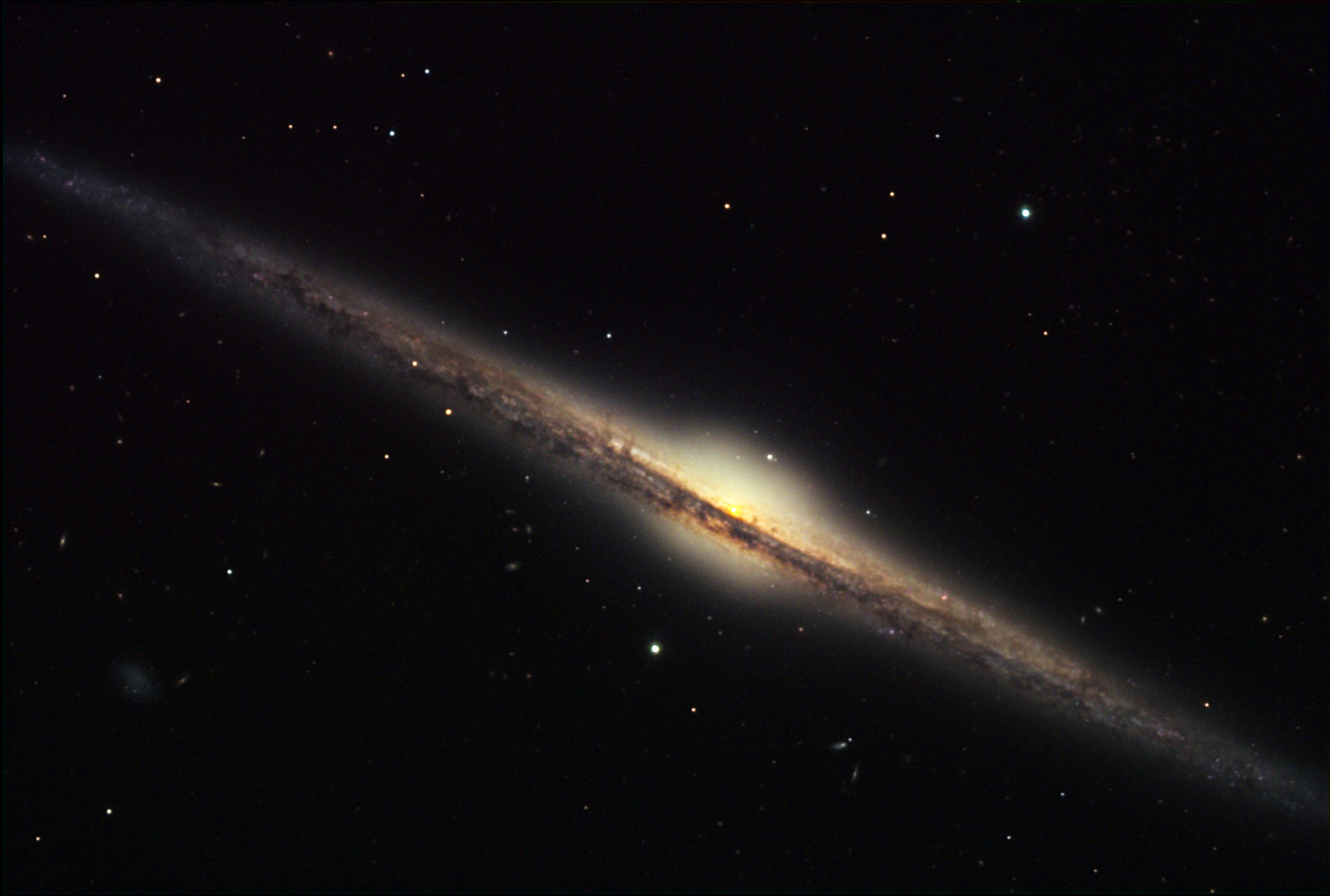}}
\caption{Images of NGC 4594 (Sombrero or Messier Object 104) galaxy, 
NGC 3031 (Messier 81) galaxy, and NGC 4565 (Needle) galaxy,  
which show the common round thin-disk structure of spiral galaxies
with small, amorphous, centrally located bulge. }  
\label{fig:fig1}
\end{figure}

Thus, it may not be unreasonable to consider a galaxy as 
an axisymmetric thin disk consisting of distributed self-gravitating mass 
in balance with a distributed centrifugal force due to 
rotation in circular orbit.
In fact, many observations and measurements of galaxies,
such as rotation curves, surface brightness, among others, 
are presented in terms of variables as functions of the 
galactocentric radius of an axisymmetric circular thin disk.
If we approximate a galaxy as a self-gravitating 
continuum of axisymmetrically distributed mass in a 
thin disk with an edge at finite radius $R_g$,
beyond which the mass density is expected to diminsh to the
inter-galactic level,
the gravitational field at a galactocentric radius $r$ can be
calculated as 
\begin{equation} \label{eq:gravitational-field}
\frac{G M_g}{R_g^2} \int_0^1 \left[\int_0^{2 \pi} 
\frac{(r - \hat{r} \cos \phi) d\phi}
{(\hat{r}^2 + r^2 - 2 \hat{r} r \cos \phi)^{3/2}}\right] 
\rho(\hat{r}) h \hat{r} d\hat{r}
\end{equation}
where all the variables are made dimensionless by 
measuring length (e.g., $r$, $\hat{r}$, $h$) 
in units of the galactic radius $R_g$, 
and mass density $\rho$ in units of
$M_g / R_g^3$ with $M_g$ denoting the total mass of the galaxy. 
Here, the disk thickness $h$ is assumed to be constant.

The centripetal acceleration of an object 
at $r$ rotating with a velocity $V(r)$ can be written as 
\begin{equation} \label{eq:centripetal-accel}
\frac{V_0^2}{R_g} \frac{V(r)^2}{r} \, \, ,
\end{equation}
where $V(r)$ is measured in units of a characteristic 
rotational velocity $V_0$.  

As shown in our previous publications 
\citep{gallo2009, gallo2010, feng2011, feng2014},
equating (\ref{eq:gravitational-field}) and (\ref{eq:centripetal-accel})
with slight algebraic arrangements yields a force-balance equation,
\begin{equation} \label{eq:force-balance}
\int_0^1 \left[
\frac{E(m)}{\hat{r} - r} - \frac{K(m)}{\hat{r} + r}
\right] 
\rho(\hat{r}) h \hat{r} d\hat{r}
+ \frac12 A V(r)^2
 = 0 \, ,
\end{equation}
where $K(m)$ and $E(m)$ denote the complete elliptic integrals of 
the first kind and second kind,
with 
\[
m \equiv \frac{4 \hat{r} r}{(\hat{r} + r)^2} \, .
\]

The dimensionless paramter $A$ in (\ref{eq:force-balance}), called 
the galactic rotation number, is defined as
\begin{equation} \label{eq:parameter-A}
A \equiv \frac{V_0^2 \, R_g}{M_g \, G} \, ,
\end{equation}
which can be determined by introducing a constraint equation
for mass conservation,
\begin{equation} \label{eq:mass-conservation}
2 \pi \int_0^1 \rho(\hat{r}) h \hat{r} d\hat{r} = 1  \, \, .
\end{equation}

\subsection{Gravitational mass}
The integral (\ref{eq:gravitational-field}) is equivalent to 
the left side of (\ref{eq:solar-mass2})
but for a distributed mass in the galactic disk, 
while (\ref{eq:centripetal-accel}) has the same physical meaning as 
the right side of (\ref{eq:solar-mass2}).
With a readily measured rotation curve\footnote{The 
measured rotation curve has been considered 
to provide the most reliable information for 
deriving the mass distribution in disk galaxies 
\citep{toomre1963, sofue2001}}---the 
orbital velocity as a function of galactocentric radius, 
$V(r)$---the mass distribution
in a galaxy can be determined
by solving for $\rho(r)$ and $A$ in
(\ref{eq:force-balance}) 
and (\ref{eq:mass-conservation}),
as elaborated in our previous publications 
\citep{gallo2009, gallo2010, feng2011, feng2014}. 
The mass determined with this method 
is fundamentally the same as that described by (\ref{eq:solar-mass2})
for determining the solar mass,
although computing $\rho(r)$ in (\ref{eq:force-balance}) 
is much more involved than calculating 
$M_{\odot}$ from (\ref{eq:solar-mass2}).
It is sometimes called the ``gravitational mass'' 
(by people who care to make the distinction)
as being derived from the gravitational field.

\begin{figure}
{\includegraphics[clip=true,scale=0.60,viewport=76 350 750 750]{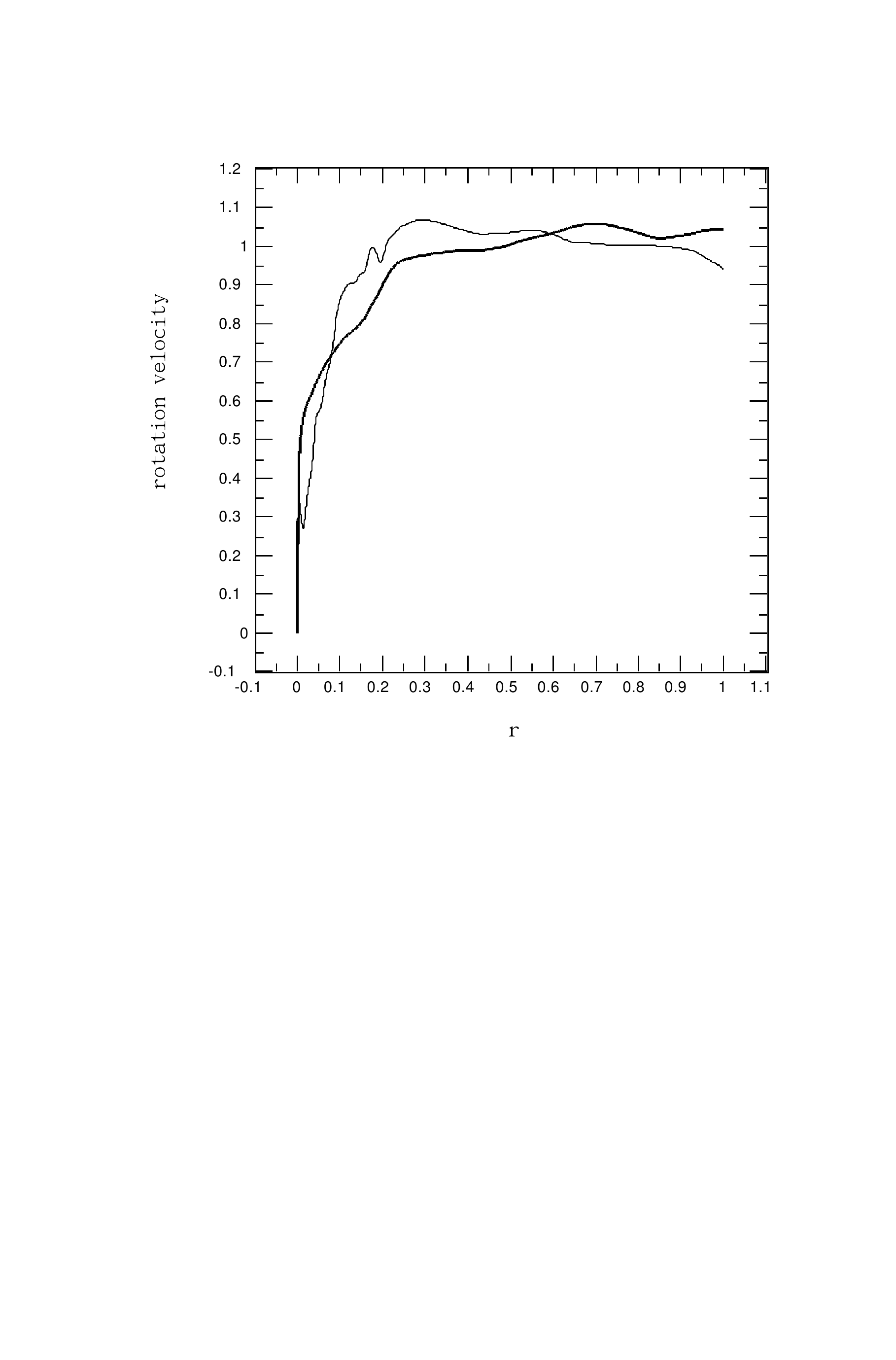}}
{\includegraphics[clip=true,scale=0.60,viewport=76 350 750 750]{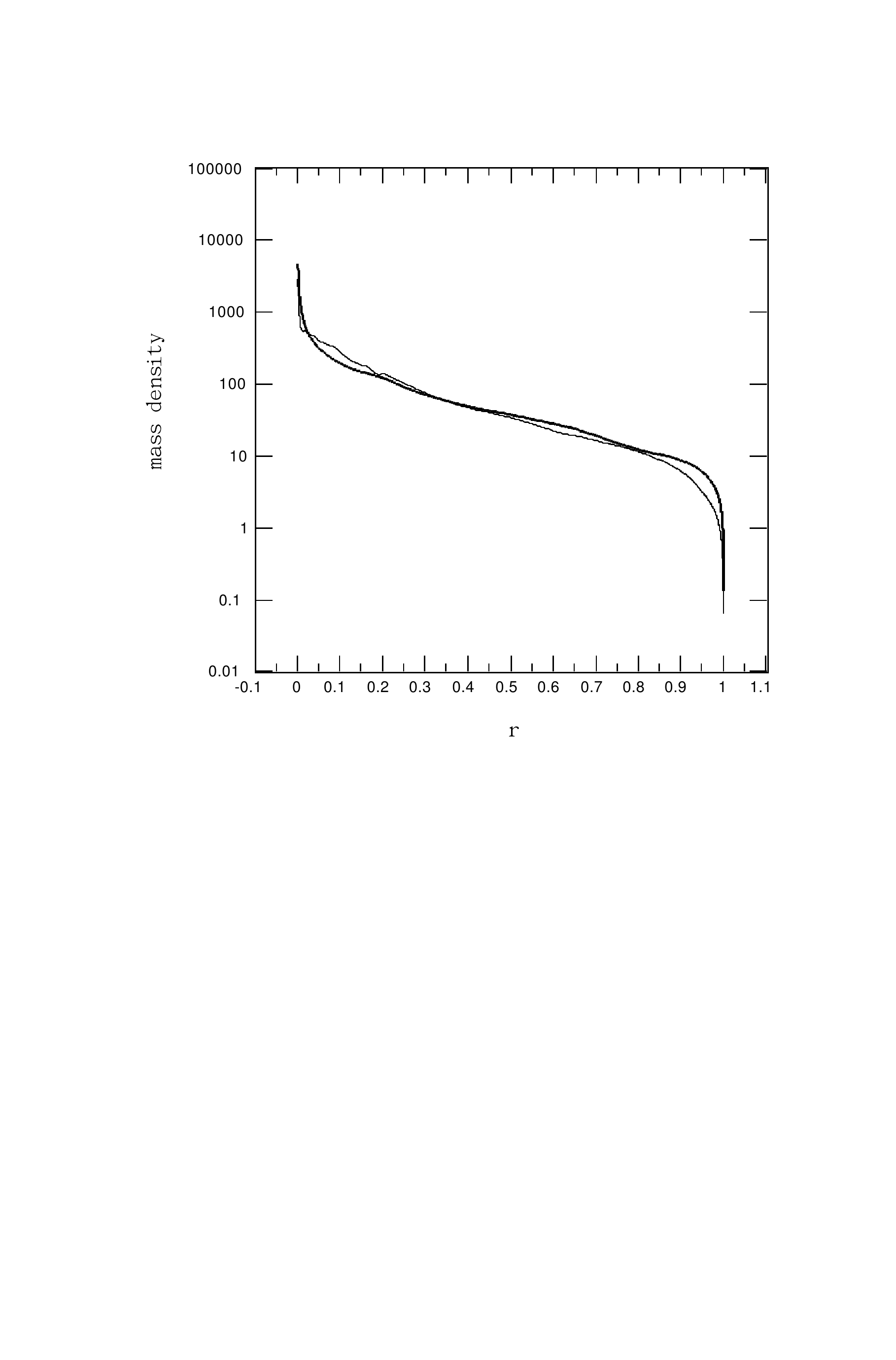}}
\caption{Profiles of measured rotation velocity $V(r)$ 
and computed mass density $\rho(r)$ for galaxies 
NGC 2403 ($R_g = 19.70$ kpc, $V_0 = 130$ km/s, 
$A = 1.4918$) and NGC 3198 
($R_g = 31.05$ kpc, $V_0 = 160$ km/s, $A = 1.6022$), 
assuming $h = 0.01$ 
\citep{feng2014}.  
A large portion of $\rho(r)$, e.g., for $0.1 \le r \le 0.9$, 
can be well approximated with a straight line in the semi-log plot,
indicating that the mass density of most spiral galaxies follows a 
common exponential law of decline as qualitatively consistent with 
typical
luminosity measurements.}  
\label{fig:fig2}
\end{figure}

With several computed examples from various types of measured rotation curves
(e.g., Figure 2),
\cite{feng2011, feng2014}
illustrated that
most surface mass density profiles $\rho(r) \, h$ 
(with the abruptly varying ends at $r = 0$ and $1$ being trimmed out)
exhibit approximately a common exponential law of decay,
qualitatively consistent with the observed 
surface brightness distributions in spiral galaxies.
As used for determining the solar mass from (\ref{eq:solar-mass2}),
Newtonian dynamics can describe the mass distribution 
in spiral galaxies self-consistently, 
according to the measured rotation curves.  
Therefore, we believe spiral galaxies described in this way
are the rotating thin-disk galaxies through the eyes of 
Newton \citep{feng2010}\footnote{The method described here 
should not be confused with the much simplified approach 
often described in textbooks and literature using 
the formula based on Keplerian dynamics, which can 
lead to erroneous results for disk galaxies \citep{feng2010}}.

\subsection{Luminous mass}
Taken at face value,
``luminous mass'' is 
the mass of an object that is luminous and one can {\em see}. 
In astronomy, observable information is carried by
``light''---electromagnetical radiation---emitted from
the {\em visible} objects.
Light can be analyzed 
to provide understanding about the emitting objects,
such as their material constituents, surface temperature,
distance, moving velocity, etc.
But to derive the amount of mass in an object from 
the light it emits does not seem to fit the common sense
based on our everyday life experiences.
In fact, objects that do not emit light
(and that can hardly be seen especially when far away) are 
quite common. 

Asteroids, for example, are rocky objects revolving around the sun that are 
too small to be called planets.  They do not emit light and 
are often hardly visible.
Even identifying asteroids has barely begun in the 21st century, because
they are not easily {\em seen} but are known to be abundant 
and of great threat to our existence.
Only a guess based on extrapolating from cratering rates on the Moon 
suggested that some two thousand asteroids big enough to imperil 
civilized existence regularly cross the Earth's orbit.
For those that had been identified (or seen), the values for 
their mass given on the NASA website in an 
Asteroid Fact Sheet \citep{asteroidfact}
are noted as only rough estimates,
which offers a hint about the level of certainty in 
measuring mass on {\em visible} objects not even too far from the Earth
in grant astronomical length scales.

Because astronomical observations rely on light,
the luminosity measurements of galaxies among other objects have been 
extensively refined and analyzed in one or more 
specified wavelength bands \citep{binney2008}.
In the late 1950s,
a systematic survey of the luminosity in spiral galaxies
led 
\cite{devaucouleurs1958}
to establish 
the universal `exponential disk' description of 
the radial surface brightness distribution 
in galactic disks.
To derive the luminous mass from measured luminosity,
the so-called mass-to-light ratio $M/L$ (in units of the solar value) 
has been used.
But to determine the value of $M/L$, the value of 
mass density (for $M$) in addition to the measured luminosity density
is needed.
Although the local mass density may in principle be derived based on
observed velocity dispersion perpendicular to the disk (in the z-direction)
for a homogeneous stellar population \citep{oort1932, oort1965}, 
the certainty 
in measuring the galactic acceleration gradient perpendicular to 
the disk plane has proved elusive \citep{faber1979}.
Even for the main-sequence stars, which are typically believed 
as having the most reliable mass-luminosity relation,
the scattering of data appears to be easily 
over orders of magnitude in the Hertzsprung-Russell diagram. 
Hence, considerable uncertainties are expected in cited values of $M/L$ 
shown in literature. 

The intrinsic unreliability in estimating masses of galaxies 
from luminosity
was pointed out by \cite{zwicky1937},
who has been credited for one of the first to use the term
``dark matter.'' 
Because of the presence of the dark matter (which doesn't emit light
but can absorb light to some unknown extent), 
masses estimated from observed luminosities are expected to be 
{\em incomplete} and can at best 
``furnish only the lowest limits'' for the actual values
\citep{zwicky1937}.
Based on Zwicky's reasoning, 
it should not be surprising to have large values of $M/L$ because
not all objects containing mass can be luminous and seen.

An examination of the methods for determining the ``$M$'' in $M/L$
(as discussed in \S 2) 
reveals the fact that $M$ of an astronomical object 
(such as the main-sequence stars) 
must come from the {\em gravitational} mass calculation 
(e.g., in the binary star systems).
Hence, the {\em luminous mass} referred to in literature
cannot be separated or obtained 
independently from the gravitational mass.
It becomes clear that 
the luminous mass, if so preferred to term it, 
can be nothing more than a portion of the gravitational mass 
that encompasses both luminous and non-luminous (or ``dark'') mass. 
Finding that the luminous mass differ from (and especially less than) 
the gravitational mass is naturally expected, and should not be surprising. 

\section{Reasoning for Dark Matter}
The reasoning for dark matter in galaxies usually
starts with the galactic rotation curve expected from
the {\em visible} or {\em luminous} matter.
Based on the apparent exponential decay of measured surface brightness 
in typical disklike galaxies and an assumption of constant $M/L$, 
\cite{freeman1970} derived an elegant analytical formula for
rotation curve $V(R)$ 
for the exponential disk
from the given surface mass density 
$\mu(R)$ 
$= \mu_0 \exp(-\alpha \, R)$, 
\begin{equation} \label{eq:freeman}
V(R)^2 =  \pi G \mu_0 \alpha R^2 
(I_0 K_0 
- I_1 K_1)  \, \, ,
\end{equation}
where $I_n$ and $K_n$ ($n = 0$ and $1$) denote modified Bessel functions
evaluated at $\alpha \, R / 2$.
The rotation curve described by (\ref{eq:freeman}) indicates that
the rotation velocity peaks at $\alpha R \approx 2.2$ 
(where $1/\alpha$ is called the radial scale length) and then declines.

If a trustworthy value of $M/L$ 
could indeed be established,
the surface mass density 
$\rho(r) \, h$ in
(\ref{eq:force-balance}) 
would then be simply obtained by multiplying the 
measured radial distribution of surface brightness distribution
with $M/L$.
Consequently, (\ref{eq:force-balance}) can be used to 
predict the rotation curve $V(r)$ from a known $\rho(r) \, h$
(of arbitrary distribution not necessarily 
described by a simple exponential function) 
as demonstrated by
\cite{feng2011}.
Here, $\rho(r) \, h$ should be considered as the luminous 
surface mass density
because it comes from the luminosity measurement.

\subsection{Flat H I rotation curves}
Until the early 1970s, 
most galactic rotation curves were measured with 
optical signals, which did not 
extend beyond the luminous regions. 
The optical rotation curves with limited extension 
appeared consistent with (\ref{eq:freeman}). 
Later, radio synthesis telescopes were constructed 
to enable measurements of the 21 cm wavelength signals emitted from 
neutral atomic hydrogen (H I) far beyond the starlight.
Against the prediction 
from the luminous mass, 
the H I rotation curves 
often do not show a decline over large radial distance.
Instead, the rotation velocity stays nearly constant
out to the limits of measurable data---which is often referred to as
the `flat' rotation curve.
It is the flat rotation curve extending far beyond 
luminous galactic disk, indicating considerable amount of mass 
existing outside the region where matter could be seen,
that has been believed to
provide the {\em compelling evidence} for a 
dark matter component to dominate the total mass of spiral galaxies. 

So, there can be some kind of ``dark'' matter that does not emit much light 
in comparison with the Sun 
(having $M/L >> 1$), to explain the difference between
gravitational mass and luminous mass.
Such dark matter was envisioned by 
those authors who initiated the term {\em dark matter} 
\citep{oort1932, zwicky1937}, which entailed no mystery 
and should not be surprising.
The neutral atomic hydrogen (H I) that enabled  
rotation curve measurement based on the 21 cm wavelength signals
exists far beyond the optical disk. 
Each hydrogen atom
carries $1.67 \times 10^{-24}$ g of mass.
If mass outside the optical disk with starlight 
is considered {\em dark},
the mass associated with H I must belong to the dark matter;
its existence suggests the fact that 
the luminous mass alone cannot account for the mass of HI 
to realistically describe the 
galactic rotation behavior,
and not all of the observable mass (including H I) can be 
derived from luminosity.
Simply put,
the luminous mass cannot (and should not) be 
the same as the gravitational mass,
as intuitively expected.
There is nothing mysterious and puzzling about the  
frequently quoted discrepancy 
between gravitational mass and luminous mass.

\subsection{Non-baryonic dark matter}
However, the current popular belief 
(of serious scientists) is 
that the dark matter inferred from the difference between 
gravitational mass and luminous mass 
is mainly non-baryonic---the kind of matter other than
the familiar protons and neutrons that make up stars and planets.
This came from that a summation of the mass of all the baryonic matter,
some of which could not be seen with the earlier optical telescopes 
but can now be detected in a myriad of new wavelengths 
with modern telescopes,
still seems to fall 
``a long way short of accounting for the effects of dark matter''
\citep{freeman2006}.
Yet most of the estimates came from different techniques 
for studying the amount of
matter at different scales with different types of 
uncertainties;
``there is always the possibility that one or all of the 
estimates could be wrong''
\citep{freeman2006}.

It is understandable that 
more sophisticated telescopes can enable seeing 
previously unseen (dark or dim) matter. 
But just seeing an object can hardly tell us the amount of mass it contains.
For example, 
the radio synthesis telescopes can detect the neutral atomic hydrogen (H I)
to enable measuring the H I rotation curves far beyond the optical disks.
What this can tell us for sure is that 
there is definitely matter (such as H I)
outside the optical disks, which was thought to be ``dark'' previously, 
now may not be dark anymore (because it can be {\em seen}).
Atomic hydrogen is known to have mass, but its $M/L$, if could be 
determined with reasonable confidence, is very unlikely 
to have the same value as that of main-sequence stars.
Although estimation of visible mass has been widely attempted 
(often with somewhat arbitrary assumptions on $M/L$),
the evaluation of associated uncertainty 
has rarely been seriously presented.
To our understanding of the current measurement techniques,
one should not be surprised by an order of magnitude uncertainty
in most of the reported data. 

Interestingly though, when \cite{oort1932} suspected
dark matter in the Galactic disk by examining vertical stellar 
motions, it was estimated as ``perhaps twice as much as was represented
by the stars of the Milky Way''---suggesting only about 
a factor of two or three, 
which could very likely be regarded as a consequence of
the unreliability in stellar mass data.
But Oort's calculation was found to be erroneous 
by later scientists based on more sophisticated analysis;
``the disk of the Milky Way---contrary to Oort's conclusion---is more
or less free of dark matter.''
However, ``Oort's `discovery' of disk dark matter would
in fact be a much more comfortable result''\footnote{probably for a desire of
rounding up the apparently needed mass to 
explain the observed 
fast motions of surrounding globular clusters 
if they were assumed not to simply be passing through} 
\citep{freeman2006}.

By computing solutions of $\rho(r)$ to (\ref{eq:force-balance})
from measured rotation curves $V(r)$, 
we found that up to the ``cut-off'' radius at $20.55$ kpc 
(beyond which there was no more detectable signal) 
a total mass of 
$1.41 \times 10^{11}$ $M_{\odot}$ 
is sufficient for supporting the Milky Way disk rotation \citep{feng2014}.
This value is very close to the oft-quoted 
``about $10^{11}$ stars'' in the Milky Way \citep{binney2008},
in view of the appreciable uncertainties in astronomical 
mass calculations. 

If we take 
$3.4 \times 10^{-20}$ kg/m$^3$
as the average gravitational mass density in the 
solar neighborhood\footnote{corresponding to  
$\sim 100$ $M_{\odot}/$pc$^2$, 
as predicted surface mass density needed at the solar galactocentric radius 
according to a measured rotation curve (with exact value in the range
of $\sim 74$ to $144$ $M_{\odot}/$pc$^2$ depending on the size 
of bulge considered in a self-gravitating disk model 
\citep{feng2014}),
and a disk thickness of $200$ pc, 
i.e., $h = 0.01$ in (\ref{eq:force-balance}), leading to 
an estimated volume mass density of $\sim 0.5$ $M_{\odot}/$pc$^3$, 
within the same order of magnitude as reported from other sources
\citep{binney2008}}, 
there should be equivalently $\sim 2 \times 10^7$ hydrogen atoms 
in a cubic meter, or equivalently $\sim 20$ hydrogen atoms per cm$^3$.
This value falls well within the reported range of estimated gas density in 
the Interstellar Medium (ISM---the matter in the space
between the stars in a galaxy, which fills interstellar space
and blends smoothly into the surrounding 
intergalactic space) of our Galaxy
\citep{ferriere2001, binney2008}.   
In terms of average mass density, 
the amount of matter in a galactic stellar system 
is extremely tenuous by terrestrial standards ($> 10^{20}$ 
atoms per cm$^3$), 
even when the stars are included. 
Beyond the solar neighborhood, the mass density in a disk model
is expected to further decrease
nearly exponentially with galactocentric radius 
according to Newtonian dynamics and measured rotation curve 
\citep{feng2011, feng2014}.
Thus, the amount of mass required to support the observed rotation curve
could be no more than that found in typical ISM.

Among the ISM, stars only form inside large complexes of 
cold molecular clouds, typically a few pc in size having 
a number density of $10^2$--$10^6$ molecules per cm$^3$ and 
a fractional volume of $< 1\%$ \citep{ferriere2001}. 
Because of the nearly exponential decrease of 
average mass density with 
galactocentric radius as shown in Figure 2,
chances for star formation are expected to diminish beyond 
a certain galactocentric distance due to lack of dense molecular clouds
where the ISM type of matter may still have relatively considerable
amount of mass.
Therefore, outside the optical disk where no more starlight can be seen,
clouds of gas in ionic, atomic, and molecular form and dust 
are likely to provide enough mass 
for explaining the observed flat rotation curve.
Only a few (baryonic) atoms per cm$^3$ in terms of the average number density
could be sufficient.
Is it really necessary to bring the non-baryonic dark matter
in for explaining the observed flat rotation curves?     

As of today,
the basis for suggesting dominant amount of non-baryonic matter 
in galaxies is neither convincingly validated
nor intuitively reasonable.

\subsection{Dark matter halo}
Moreover, the current popular model of a spiral galaxy 
consists of a decomposition of a bulge, a disk, 
and a dark matter halo 
\citep{faber1979, sofue2001, freeman2006, binney2008}.
The bulge---a small, amorphous, centrally located stellar system---and 
the disk with well-defined rotation curve 
are evidenced by optical observations (cf. Figure 1).

But the dark matter halo has no direct observational basis, 
with rather vague, brief explanations, if any, being provided in books
and literature.
According to several authors
\citep{faber1979, sofue2001, freeman2006, binney2008}.
the concept of dark matter halo sounded like it was
coming from the N-body simulation of 
\cite{ostriker1973},
arguing that the most plausible way to stablilize the Galaxy
against the bar instability 
was to add a massive dark halo.
The dark halo was believed to provide 
part of the equilibrium gravitational field,
thereby reducing the required disk mass and the destabilizing effect of 
the disk's self-gravity.
But there are other more recent N-body simulations showing that
a disk galaxy with an almost flat roation curve can be stabilized by 
dense centers 
without the dark matter halo \citep{sellwood2001},
reversing the argument for requiring a massive halo to
stabilize the disk.
It was also suggested that halos are not very efficient for
stabilizing the disk as compared to bulges \citep{kalnajs1987}.

Yet still, the dark matter halos have been considered as 
an indispensable component of galaxies,
despite that many authors tend to believe
negligible mass contribution from the dark matter halo 
in the luminous disk region.
This led to the so-called maximum disk hypothesis 
\citep{vanalbada1986},
which assumes the mass of the luminous disk to be as large as possible
to produce the galactic rotation curve,
with the mass of the dark matter halo dominant in the outer region beyond the 
visible edge.
But there is no clear explanation on why the dark matter halo 
takes a spherical shape and stays only in the outer region,
except for the modeling convenience by assuming it 
as an isothermal sphere \citep{carignan1985}.  
Even to this day, 
``the shape of dark matter halos remains a mystery''
\citep{freeman2006}.
Thus, the assumption of a dark matter halo containing substantial 
amount of mass 
lacks supporting evidence and logical rigor\footnote{From the modeling 
point of view, a pure thin disk based on a given rotation curve
can yield a uniquely determined surface mass density distribution
in a spiral galaxy;
adding a spherical bulge or halo inevitably induces the requirements of 
known spherical mass distribution which, 
to our knowledge, can 
only come from debatable assumptions \citep{feng2014}}.

\subsection{Dark matter in bigger pictures}
Outside the rotating Galactic disk, 
some randomly moving stars form a nearly spherical stellar halo.
Those halo stars are ``fast-moving, energetic stars,
buzzing around the Galaxy like a swarm of bees'' \citep{freeman2006}.
Our galaxy also contains about $150$ 
globular clusters---spherical collections of 
$10^4$--$10^6$ stars \citep{binney2008}.
Some of those halo stars or globular clusters may move at 
velocities so high that they exceed the 
estimated escape velocity of the Galaxy. 
If those fast moving objects were bound to the Galaxy
by gravitation,
much more Galactic mass than that to support the rotation curve 
would be needed.
Thus, the idea of invisible dark matter could become 
entertaining.
However, observations of the entire orbits of those randomly
moving objects, which individually may take 
hundreds of million years to complete,
can be extremely challenging. 
Without complete orbital knowledge about those halo stars,
snapshots of
their transient motions can hardly be taken as an observational evidence 
for dark matter. 

In a larger scale, galaxies (or `nebulae') appear to congregate in groups,
called galaxy clusters,
instead of being randomly distributed 
non-interacting loners in space.
\cite{zwicky1937} investigated 
the Coma cluster by applying the virial theorem 
with a tentative hypothesis of ``statistically stationary system.''
From the average velocity of observed galaxy motions,
Zwicky estimated the kinetic energy of the system 
and found that the mass needed
for the gravitational potential energy 
to prevent those galaxies from flying apart is about 
$4.5 \times 10^{13}$ $M_{\odot}$.
With about $1000$ galaxies in the Coma cluster,
this led to the average mass of each galaxy to be about 
$4.5 \times 10^{10}$ $M_{\odot}$ \citep[about one third of 
that of the Milky Way according to][]{feng2014},
which was ``somewhat unexpected'' when compared to 
$8.5 \times 10^{7}$ $M_{\odot}$ as suggested by the luminosity data.
Therefrom, Zwicky has been credited as one of the first scientists for 
discovering the {\em dark matter} or missing mass.
However, Zwicky in the same article was quite critical about the 
intrinsic unreliability in estimating galactic mass from 
its luminosity (which was rarely mentioned in literature
when talking about his first usage of the term dark matter,
when he considered objects ``in the form of 
cool and cold stars, macroscopic and microscopic solid bodies 
and gases.'')

From a historical perspective, Zwicky was an admirably  
intelligent scientist who made many important contributions
not only in astonomy but also in a wide range of 
other disciplines.  He was one of the first to study
galaxy clusters, through which he suggested 
serious discrepancy between gravitational mass and 
luminous mass which implied considerable amount of unseen or 
dark matter. 
However, using Zwicky's virial theorem calculations 
as the evidence for dark matter (especially the 
non-baryonic dark matter) 
seems to be too much of a stretch,
in view of what we know about the level of uncertainties 
in estimating the luminous mass and counting the baryonic mass. 
In this regard, we should keep in mind that 
the volume of space between the stars and galaxies 
is tremendous; 
just miscounting a matter of a few atoms in a cubic centimeter 
in the interstellar medium, 
especially in the outer regions where the volume 
increases rapidly  
with the galactocentric radius, 
may easily lead to an order of magnitude difference 
in the estimated galactic mass.

When it comes to cosmology,
the Big Bang Theory seems to be the present-day standard model of 
the origin of the Universe. 
It suggests that dark matter accounts for 
about $27\%$ 
while ordinary (baryonic) matter 
accounts for only $4.9\%$ of the mass-energy content of 
the observable universe,
with the remainder being attributable to {\em dark energy}.
Inasmuch as its acceptance as the current standard model,
the Big Bang Theory has not been fully verified by 
the existing observational evidence and 
remains as a hypothetical theory.
Alternative models have been proposed for our understanding of 
the Universe as a whole \citep{corredoira2014}. 
In fact, baryonic matter exists everywhere. 
Even in the vast expanses between the galaxies
where it had always been assumed more or less empty,
an entirely unsuspected source of stellar baryonic matter 
has recently been discovered \citep{freeman2006}.

\section{Conclusions}
When astronomers and astrophysicists found 
a difference or discrepancy between the gravitational mass
and luminous mass in galaxies,
they were puzzled and resorted to a mysterious matter 
that could not be seen---the {\em dark matter}.
Some scientists would even consider such discrepancy
as the {\em compelling} evidence for dark matter,
as if the confidence level were quite high.
However, little details could be found in 
quantifying the cascade of uncertainty in  
determining the luminous mass and gravitational mass,
which is normally crucial when analyzing 
the difference between two quantities and 
establishing the confidence level thereof. 

Estimating mass of an astronomical (or celestial) object far away is 
understandably challenging with many limitations in methods and tools.
The consequence of the difficulties involved 
in measuring the distance of a star
or a galaxy is a considerable level of uncertainty.
Measuring the luminosity of a star depends on its distance
which undoubtedly cascades its uncertainty
to the luminosity calculation.
Direct measurement of the mass of a star seems to be limited 
to the (eclipsing) binary star systems in the solar neighborhood, 
which depends on not only its distance 
but also the separation between the two stars. 
With the available data of mass and luminosity,
an empirical mass-to-light ratio of a star can be obtained,
which is then applied to 
estimating masses of other stars of the same type (i.e.,
sharing the similar color or surface temperature) whose 
direct mass measurement is intractable.
Thus, the ``luminous mass'' determined from 
the luminosity and mass-to-light ratio can be at best 
a rather crude estimate,
considering the aggregation and cascade of uncertainty involved
in many converting steps based on assumptions not thoroughly verified.
Of course, for each individual galaxy, derivation of its gravitational mass 
from the measured rotation curve 
also involves uncertainties 
associated with the measurements of stellar motion velocity and distance.
But many observed spiral galaxies exhibit quite similar 
structural configurations and rotation characteristics,
which at least offer some level of statistical confidence.

In view of the fact that all mass measurements of the 
astronomical objects are based on gravitational force, 
even that for estimating the mass-to-light ratio of a star
which is in turn used to determine the luminous mass,
the gravitational mass (which comes directly from gravitational 
effect measurement) is intuitively expected to be 
much more reliable than the luminous mass.
In fact, the measured rotation curves 
have been considered to provide the most reliable means for 
determining the distribution of gravitational mass in 
spiral galaxies \citep{toomre1963, sofue2001}.
By closely examining the methods for evaluating the luminous mass,
one can quickly realize that the so-called luminous mass
can at best represent a portion of the gravitational mass,
and it is not supposed to match the gravitational mass.
Therefore, it is natural to find less luminous mass than 
gravitational mass.

In general, the stellar systems such as galaxies are 
extremely tenuous in terms of average mass density.
For example, our Newtonian dynamics model based on measured rotation curve
predicted that the average (gravitational) mass density in 
the solar neighborhood is around 
$0.5$ $M_{\odot}/$pc$^3$ \citep{feng2014}  
(or $10^{-20}$ kg$/$m$^3$).
This corresponds to about $20$ hydrogen atoms per cm$^3$,
well within the reported range of 
estimated mass density in the interstellar medium.
Given the vast volume in a typical galaxy, 
a slight misscounting of matter due to observational limitations 
can cause huge variations in evaluation of the galactic mass. 
With this kind of perspective in mind, 
one would natually wonder whether
quantitative evaluation of the discrepancy between 
luminous mass and gravitational mass,
especially for establishing the {\em compelling} evidence of missing mass,
can really bear any meaningful fruits.
 
Actually the gravitational mass,
as determined from a thin disk model 
with a given rotation curve according to Newtonian dynamics,
can be quite reasonable when compared with the star counts 
(which relates to the luminous mass).
For example, the Milky Way rotation curve with a cut-off radius 
at $20.55$ kpc leads to a total mass of $1.41 \times 10^{11}$ $M_{\odot}$, 
which is fairly close to the star counts of about $100$ billion \citep{feng2014},
in view of the level of uncertainty in the stellar mass measurements. 
As far as the observed rotation curves are concerned,
the corresponding gravitational mass distribution
could be reasonably consistent with  
the distribution of luminosity if the assumption of a constant 
mass-to-light ratio is abandoned.
And there is no concrete reason for having a constant mass-to-light ratio
across the entire galaxy, except probably for the convenience of 
model calculations.
If so, the existence of dark matter 
in galaxies
may become baseless.

In our opinion, a galaxy with star type distribution varying 
from its center to periphery may have 
a mass-to-light ratio varying accordingly depending on 
the galactocentric radius.
Based on our thin-disk model computations from measured rotation curves, 
most galaxies have a typical mass density profile with 
a peak value at the galactic center falling rapidly
within $\sim 5\%$ of the cut-off radius, and then declining 
nearly exponentially toward the edge (e.g., figure 2).
Because the radial scale length for the exponentially declining portion 
is usually larger than that of the luminosity one,
the mass-to-light ratio should increase with galactocenteric radius,
corresponding to cooler and ``darker'' matter toward the peripery of 
the galactic disk as consistent with the edge-on view images 
of spiral galaxies
(cf. figure 1). 
The predicted mass density in the Galactic disk is well within the 
range of that observed interstellar medium,
and therefore can be considered reasonable. 
This leads us to believe that ordinary baryonic matter 
can be sufficient for supporting the observed galactic rotation curves;
speculation of large amounts of non-baryonic matter may be
based on an ill-conceived discrepancy between gravitational mass and 
luminous mass which appears to be unjustified. 
 
If we follow the same vein of thought, 
serious shortcomings of 
the arguments for dominant amounts of dark matter 
needed to hold a galaxy clusters together, and the like, 
are readily revealed. 
Our logical analysis presented here 
demonstrates a philosophical truth that 
one should not place too much faith
in a reported result, no matter how eminent the
scientist who presents it.
Without doubt and skepticism,
science cannot thrive and will stagnate.


\begin{thebibliography}{1}

\bibitem[Bennett et al.(2007)]{bennett2007}
Bennett, J., Donahue, M., Schneider, N., and Voit, M. 2007
{\em Cosmic Perspective: Stars, Galaxies, and Cosmology}, 
Addison and Wesley, Reading

\bibitem[Binney and Tremaine(2008)]{binney2008}
Binney, J. and Tremaine, S. 2008 
{\em Galactic Dynamics}, Princeton University Press, Princeton

\bibitem[Carignan and Freeman(1985)]{carignan1985}
Carignan, C. and Freeman, K. C. 1985 
Basic parameters of dark halos in late-type sspirals.
{\em Astrophys. J.} {\bf 294}, 494--501

\bibitem[de Vaucouleurs(1958)]{devaucouleurs1958}
de Vaucouleurs, G. 1958 
Photoelectric photometry of the Andromeda nebula in the UBV system.
{\em Astrophys. J.} {\bf 128}, 465--488

\bibitem[Faber and Gallagher(1979)]{faber1979}
Faber, S. M. and Gallagher, J. S. 1979 
Masses and mass-to-light ratios of galaxies.
{\em Ann. Rev. Astron. Astrophys.} {\bf 17}, 135--187

\bibitem[Feng and Gallo(2010)]{feng2010}
Feng, J.\ Q. and Gallo, C. F. 2010 
Rotating thin-disk galaxies through the eyes of Newton. 
{\em arXiv:1007:3778} 

\bibitem[Feng and Gallo(2011)]{feng2011}
Feng, J.\ Q. and Gallo, C. F. 2011 
Modeling the Newtonian dynamics for rotation curve analysis
of thin-disk galaxies.
{\em Res. Astron. Astrophys.} {\bf 11}, 1429--1448

\bibitem[Feng and Gallo(2014)]{feng2014}
Feng, J.\ Q. and Gallo, C. F. 2014 
Mass distribution in rotating thin-disk galaxies according to
Newtonian dynamics.
{\em Galaxies} {\bf 2}, 199--222

\bibitem[Ferriere(2001)]{ferriere2001}
Ferriere, K. M. 2001 
The interstellar environment of our galaxy..
{\em Rev. Mod. Phys.} {\bf 73(4)}, 1031--1066

\bibitem[Freeman(1970)]{freeman1970}
Freeman, K. C. 1970 
On the disks of spiral and S0 galaxies.
{\em Astrophys. J.} {\bf 160}, 811--830

\bibitem[Freeman and McNamara(2006)]{freeman2006}
Freeman, K. C. and McNamara, G. 2006 
{\em In Search of Dark Matter}, Springer, Praxis Publishing, Chichester, UK 

\bibitem[Gallo and Feng(2009)]{gallo2009}
Gallo, C. F. and Feng, J.\ Q. 2009 
A thin-disk gravitational model for galactic rotation.
In {\em 2nd Crisis in Cosmology Conference,
ASP Conf. Ser. 413} 
ed. F. Potter, San Francisco, ASP. 289--303

\bibitem[Gallo and Feng(2010)]{gallo2010}
Gallo, C. F. and Feng, J.\ Q. 2010 
Galactic rotation described by a thin-disk gravitational model
without dark matter.
{\em J. Cosmology} {\bf 6}, 1373--1780

\bibitem[Kalnajs(1987)]{kalnajs1987}
Kalnajs, A. J. 1978 
Halos and disk stability.
In {\em Dark Matter in the Universe; Proceedings of the IAU 
Symposium} D. Reidel Publishing, 289--296

\bibitem[Lopez-Corredoira(2014)]{corredoira2014}
Lopez-Corredoira, M. 2014
Non-standard models and the sociology of cosmology.
{\em Stud. Hist. 
Philos. Mod. Phys.} {\bf 46}, 86--96
(arXiv:1311.6324)

\bibitem[NASA website()]{asteroidfact}
NASA website, http://nssdc.gsfc.nasa.gov/planetary/ factsheet/asteroidfact.html

\bibitem[Oort(1932)]{oort1932}
Oort, J. H. 1932 
The force exerted by the stellar system in the direction
perpendicular to the galactic plane and some related problems.
{\em Bull. Astr. Inst. Netherlands} {\bf 6(238)}, 249--287

\bibitem[Oort(1965)]{oort1965}
Oort, J. H. 1965 
Stellar dynamics.
In {\em Galactic Structure} ed. A Blaauw, M. Schmidt, pp 455--512
Chicago University Press

\bibitem[Ostriker and Peebles(1973)]{ostriker1973}
Ostriker, J. P. and Peebles, P. J. E.  1973 
A numerical study of the stability of flatterned galaxies: or,
can gold galaxies survive?
{\em Astrophys. J.} {\bf 186}, 467--480

\bibitem[Sellwood and Evans(2001)]{sellwood2001}
Sellwood, J. A. and Evans, N. W. 2001 
The stability of disks in cusped potentials.
{\em Astrophys. J.} {\bf 546}, 176--188

\bibitem[Sofue and Rubin(2001)]{sofue2001}
Sofue, Y. and Rubin, V. C. 2001 
Rotation curves of spiral galaxies.
{\em Ann. Rev. Astron. Astrophys.} {\bf 39}, 137--174

\bibitem[Toomre(1963)]{toomre1963}
Toomre, A. 1963 
On the distribution of matter within highly flattened galaxies.
{\em Astrophys. J.} {\bf 138}, 385--392

\bibitem[Trimble(1987)]{trimble1987}
Trimble, V. 1987 Existence and nature of dark matter in the universe.
{\em Ann. Rev. Astron. Astrophys.} {\bf 25}, 425--472 

\bibitem[van Albada and Sancisi(1986)]{vanalbada1986}
van Albada, G. D. and Sancisi, R. 1986 
Dark matter in spiral galaxies.
{\em Phil. Trans. R. Soc. Lond.} {\bf A320}, 447--464

\bibitem[Zwicky(1937)]{zwicky1937}
Zwicky, F. 1937 
On the masses of nebulae and clusters of nebulae.
{\em Astrophys. J.} {\bf 86(3)}, 217--246

\end{thebibliography}
\end{document}